\documentclass[aps,prc,twocolumn,superscriptaddress,groupedaddress]{revtex4}  % for review and submission
\usepackage{graphicx}  % needed for figures
\usepackage{dcolumn}   % needed for some tables
\usepackage{bm}        % for math
\usepackage{amssymb}   % for math
\usepackage{amsmath}

\begin{document}

\title{Measuring the surface thickness of the weak charge density of nuclei}
\author{Brendan Reed}
\affiliation{Center for the Exploration of Energy and Matter and Department of Physics, Indiana University, Bloomington, IN 47405, USA}                             
\author{Z. Jaffe}
\affiliation{Center for the Exploration of Energy and Matter and Department of Physics, Indiana University, Bloomington, IN 47405, USA}                             
\author{C. J. Horowitz}       
\affiliation{Center for the Exploration of Energy and Matter and Department of Physics, Indiana University, Bloomington, IN 47405, USA}                             
\author{C.~Sfienti}
\address{Institut f\"ur Kernphysik, Johannes Gutenberg-Universit\"at Mainz, D-55099 Mainz, Germany}

\date{\today}

\begin{abstract}
\noindent
The present PREX-II and CREX experiments are measuring the rms radius of the weak charge density of $^{208}$Pb and $^{48}$Ca.  We discuss the feasibility of a new parity violating electron scattering experiment to measure the surface thickness of the weak charge density of a heavy nucleus.  Once PREX-II and CREX have constrained weak radii, an additional parity violating measurement at a momentum transfer near 0.76 fm$^{-1}$ for $^{208}$Pb or 1.28 fm$^{-1}$ for $^{48}$Ca can determine the surface thickness.  

\end{abstract}

%\pacs{25.30.Bf, 27.40.+z, 21.10.Gv, 21.10.Ft}
\maketitle

%\section{\label{sec:level1}First-level heading}
% sections are not used for PRL papers
\section{Introduction}

Where are the protons located in an atomic nucleus?  Historically, charge densities from elastic electron scattering have provided accurate and model independent information \cite{1}.  These densities are, quite literally, our current picture of the nucleus and they have had an enormous impact.  They have helped reveal sizes, surface thicknesses, shell structure, and saturation density of nuclei.    

An equally important but much more challenging question is where are the neutrons located in an atomic nucleus? Very fundamental nuclear structure information could be extracted if we also had access to accurate neutron densities.  For example, knowing both the proton and the neutron densities would provide constraints on the isovector channel of the nuclear effective interaction, which is essential for the structure of very neutron rich exotic nuclei.  Because of nuclear saturation, we expect the average interior baryon density of $^{208}$Pb to be very closely related to the saturation density of nuclear matter $\rho_0$.  Since we already know the charge and proton densities with high precision, determining the interior neutron density of $^{208}$Pb would allow new insight into nuclear saturation and the exact value of $\rho_0$ \cite{nuclearsaturation}.

However, compared to charge densities, our present knowledge of neutron densities is relatively poor and model dependent.  Often neutron densities are determined with strongly interacting probes \cite{2} such as antiprotons \cite{3,4}, proton elastic scattering \cite{5}, heavy ion collisions \cite{7}, pion elastic scattering \cite{8}, and coherent pion photo-production \cite{9}.  Here one typically measures the convolution of the neutron density with an effective strong interaction range for the probe.  Uncertainties in this range, from complexities of the strong interactions, can introduce significant systematic errors in the extracted neutron densities (see Ref. \cite{tt} for a recent review of neutron skin measurements).

It is also possible to measure neutron densities, or equivalently weak charge densities, with electro-weak interactions, by using neutrino-nucleus coherent scattering \cite{coherent,coherentAr,10,11} or parity violating electron scattering \cite{13,20}.   Since the weak charge of a neutron is much larger than that of a proton, the weak charge density of a nucleus is very closely related to its neutron density.  Compared to strongly interacting probes, parity violation provides a clean and model-independent way to determine the weak charge density and it is likely affected by much smaller strong interaction uncertainties.  In the last few decades, great theoretical \cite{12,13,14,15,16,17,18}  and experimental \cite{19,20} efforts have been made to improve parity violating electron scattering experiments.  At Jefferson laboratory, the radius of the weak charge density of $^{208}$Pb was first measured in the PREX experiment \cite{20,22}, and has been recently measured with higher accuracy in the PREX-II  experiment \cite{PREXII}, while the weak radius of $^{48}$Ca is being measured in the CREX experiment \cite{CREX}.

It is a slight misnomer to say PREX-II and CREX are directly measuring weak radii.  Strictly speaking the radius is defined by the derivative of a weak form factor (see below) in the limit of the momentum transfer going to zero.  However for practical reasons, PREX-II and CREX do not measure at zero momentum transfer but at small finite momentum transfers.  Therefore PREX-II and CREX depend on not just the weak radius but also, to some degree, on the surface thickness of the weak density.   In this paper we quantify this dependence and explore how the surface thickness can be determined by measuring the parity violating asymmetry at a second, somewhat higher, momentum transfer.  
% tt I would phrase this part (below) differently (see the end of the paragraph)
%Such an experiment would provide a next step after PREX-II and CREX.  In previous work we showed how the entire weak density distribution $\rho_W(r)$, and not just the radius and surface thickness, can be determined by measuring parity violation at several different momentum transfers \cite{fullweak}.   This is perhaps the ultimate goal for parity violating electron scattering experiments on heavy nuclei.  Nevertheless, a measurement of the surface thickness would provide a very useful result that is intermediate between the present PREX-II and CREX experiments and determining the full $\rho_W(r)$.        
%
Evidently such an experiment would provide the next step after PREX-II and CREX.  The ultimate goal for parity violation experiment on heavy nuclei will be, as we have shown in previous work \cite{fullweak} the determination of the entire weak density distribution $\rho_W(r)$ by measuring parity violation at several different momentum transfers.   

Our formalism for describing parity violating electron scattering and how this depends on properties of the weak density including the surface thickness is presented in Sec. \ref{sec.formalism}.  In Sec. \ref{sec.results} we present results for the sensitivity of the PREX-II and CREX experiments to the surface thickness.  Next, in Sec. \ref{sec.additional} we explore the feasibility of a new experiment to measure the surface thickness of the weak density of $^{208}$Pb or $^{48}$Ca.  We conclude in Sec. \ref{sec.conclusion}. 

%to determine the cross-section for longitudinally polarized electrons scattered from $^{48}$Ca and the parity violating asymmetry $A_{pv}$ is presented in Section II.  In Section III we motivate measuring the full radial dependence of the weak charge density in $^{48}$Ca and discuss the large information that it contains.  In Section IV we illustrate our formalism with an example experiment and calculate the resulting statistical errors.  The resulting weak density can be directly compared to modern microscopic calculations of the ground state structure of $^{48}$Ca using Chiral effective field theory interactions \cite{cc48Ca}.  We conclude in Section V that it is feasible to measure the full weak density distribution of $^{48}$Ca.  However this may be much harder for a significantly heavier nucleus such as $^{208}$Pb because more Fourier Bessel coefficients likely will be needed.    

\section{Formalism}
\label{sec.formalism}
The parity violating asymmetry for longitudinally polarized electrons scattering from a spin zero nucleus, $A_{pv}$, is the key observable which is very sensitive to the weak charge distribution.  The close relationship between $A_{pv}$ and the weak charge density $\rho_W(r)$ can be readily seen in the Born approximation, 
\begin{equation} 
A_{pv}\equiv\frac{d\sigma/d\Omega_R - d\sigma/d\Omega_L}{d\sigma/d\Omega_R + d\sigma/d\Omega_L}\approx-\frac{G_Fq^2}{4\pi\alpha\sqrt{2}}\frac{Q_WF_W(q^2)}{ZF_{ch}(q^2)}.
\label{eq:1}
\end{equation}
Here $d\sigma/d\Omega_R$ ($d\sigma/d\Omega_L$) is the cross section for positive (negative) helicity electrons, $G_F$ is the Fermi constant, $q$ the momentum transfer, $\alpha$ the fine structure constant, and $F_W(q^2)$ and $F_{ch}(q^2)$ are the weak and charge form factors respectively, 
\begin{equation}
F_W(q^2)=\frac{1}{Q_W}\int d^3r j_0(qr) \rho_{W}(r)
\label{eq:2}
\end{equation}
\begin{equation}
F_{ch}(q^2)=\frac{1}{Z}\int d^3r j_0(qr)\rho_{ch}(r).
\label{eq:3}
\end{equation}
These are normalized so that $F_W(q\rightarrow0)=F_{ch}(q\rightarrow0)=1$.  The charge density is $\rho_{ch}(r)$ and $Z=\int d^3r \rho_{ch}(r)$ is the total charge.  Finally, the weak charge density $\rho_W(r)$ and the total weak charge $Q_W=\int d^3r \rho_W(r)$ are discussed below.  

\begin{figure}[hbt]
\includegraphics[scale=0.3]{{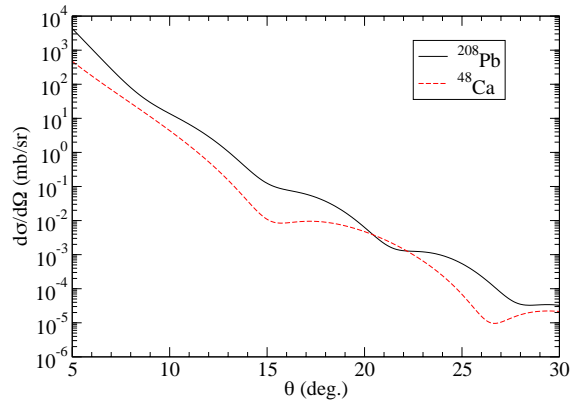}}
 \caption{\label{fig:1} Differential cross section including Coulomb distortions for 855 MeV electrons elastically scattered from $^{208}$Pb (solid black line) and $^{48}$Ca (dashed red line) versus scattering angle.  }
 \end{figure}

\begin{figure}[hbt]
\includegraphics[scale=0.3]{{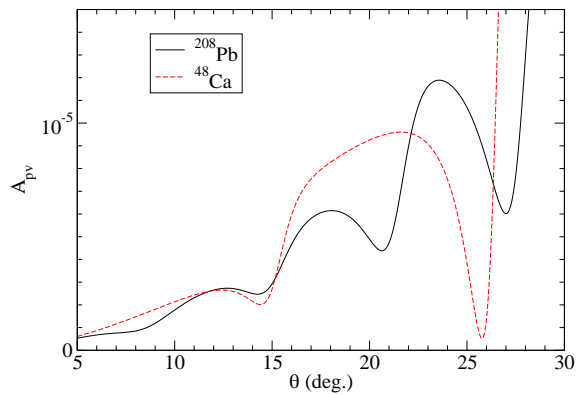}}
 \caption{\label{fig:1b} Parity violating asymmetry including Coulomb distortions for 855 MeV electrons elastically scattered from $^{208}$Pb (solid black line) and $^{48}$Ca (dashed red line) versus scattering angle.  Symmetrized Fermi weak charge densities are used (see text).}
 \end{figure}

The elastic cross-section in the plane wave Born approximation is,
\begin{equation}
\frac{d\sigma}{d\Omega}=\frac{Z^2\alpha^2\cos^2 (\frac{\theta}{2})}{4E^2\sin^4 (\frac{\theta}{2})}\bigl|F_{ch}(q^2)\bigr|^2,
\label{eq:4}
\end{equation} 
with $\theta$ the scattering angle.  However, for a heavy nucleus, Coulomb-distortion effects must be included.  These correct both Eqs. \ref{eq:1} and \ref{eq:4} and can be included exactly by numerically solving the Dirac equation for an electron moving in the coulomb and axial vector weak potentials \cite{12}.  Figure \ref{fig:1} shows the cross section and Fig. \ref{fig:1b} the parity violating asymmetry $A_{pv}$ for 855 MeV electrons scattering from $^{208}$Pb or $^{48}$Ca.    

%compare the plane-wave cross-section, Eq. \ref{eq:4}, to the cross section including Coulomb-distortion effects, see for example \cite{17}. Coulomb distortions are seen to fill in the diffraction minima.  However away from these minima distortion effects on the cross section are relatively small.  The cross section calculated with the charge density from a relativistic mean field model using the FSU-gold interaction \cite{FSUGold}, see Fig. \ref{fig:2}, agrees well with the experimental charge density except at the largest angles.

Parity violating experiments directly depend on the weak density $\rho_W(r)$.  However, theoretical calculations often determine $\rho_W(r)$ by folding single nucleon weak form factors with point proton $\rho_p(r)$ and neutron $\rho_n(r)$ densities.  For completeness, we review this procedure here.  

If one neglects spin-orbit currents that are discussed in Ref. \cite{spin_orbit}, and other meson exchange currents \cite{big_paper} one can write,
\begin{eqnarray}
\rho_{W}(r)&=&\int d^3r' \bigl\{4G_n^Z(\lvert{r-r'}\rvert)\rho_n(r)\nonumber\\
&+&4G_p^Z(\lvert{r-r'}\rvert)\rho_p(r)\bigr\}\, .
\label{eq:7}
\end{eqnarray}
Here $G_n^Z(r)$ and $G_p^Z(r)$ are the Fourier transforms of the neutron and proton single nucleon weak form factors \cite{weakpr},  
\begin{equation}
4G_n^Z(r)=Q_nG_E^p(r)+Q_pG_E^n(r)-G_E^s(r),
\end{equation}
\begin{equation}
4G_p^Z(r)=Q_pG_E^p(r)+Q_nG_E^n(r)-G_E^s(r),
\end{equation}
where $G_E^p(r)$ and $G_E^n(r)$ are Fourier transforms of the proton and neutron electric form factors.  %They are normalized
%\begin{equation}
%\int d^3r G_E^p(r)=1, \ \ \ \ \ \ \ \ \int d^3r G_E^n(r)=0.
%\end{equation}
Finally $G_E^s(r)$ describes strange quark contributions to the nucleon electric form factor \cite{strange1,strange2,strange3,strange4}.  This is measured to be small so we assume $G_E^s(r)\approx 0$.

The weak form factors are normalized,
%\begin{equation}
$\int d^3r\, 4G_n^Z(r)= Q_n$, and $\int d^3r\, 4G_p^Z(r)=Q_p$.
%\end{equation}
The weak charge of the neutron is $Q_n=-1$, while the weak charge of the proton is $Q_p\approx 0.05$, to lowest order.  Including radiative corrections \cite{rad1,rad2} one has,
%\begin{equation}
$Q_n$=-0.9878, and $Q_p$=0.0721.
%\end{equation}
Finally, the total weak charge of a nucleus,
\begin{equation}
Q_W=\int d^3r \rho_W(r)=NQ_n+ZQ_p\,.
\label{eq.qw}
\end{equation}
%is listed in Table \ref{tab:table1}.
 %Further nuclear radiative corrections, for example from $\gamma-Z$ box diagrams \cite{gzbox1,gzbox2}, are not expected to be important compared to these large values of $Q_W$.

\begin{table}[hbt]
\begin{ruledtabular}
\begin{tabular}{cccc}
 Nucleus &  $c$ (fm)   & $a$ (fm) & $Q_W$ \\
\hline
 $^{48}$Ca  &  3.99595   & 0.51540 &  -26.2164 \\
$^{208}$Pb & 6.81507 & 0.61395 & -118.551 \\
\end{tabular}
\end{ruledtabular}
\caption{\label{tab:table1} Fermi function fits of the radius $c$ and surface thickness $a$ parameters, see Eq. \ref{eq.2pf}, to weak charge densities predicted by the FSU Gold relativistic mean field interaction \cite{FSUGold}.  Also listed is the total weak charge $Q_W$, see Eq. \ref{eq.qw}.}
\end{table}

\begin{figure}[hbt]
 \includegraphics[scale=0.3]{{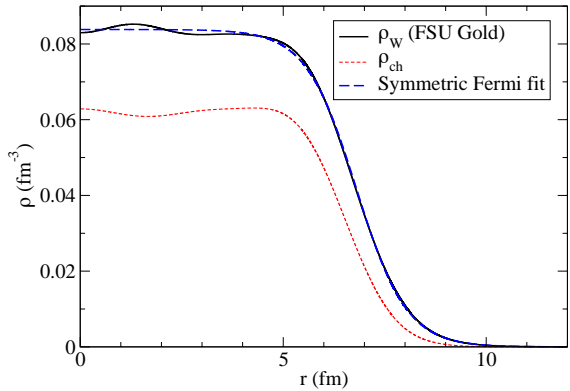}}
 \caption{\label{fig:2} Minus the weak charge density -$\rho_W(r)$ of $^{208}$Pb versus radius $r$.  The solid black line is from the FSU Gold relativistic mean field model \cite{FSUGold} while the the long dashed blue line shows a Fermi function fit (Eq. \ref{eq.2pf}).  The short dashed red line shows the experimental charge density $\rho_{ch}(r)$ from Ref. \cite{1}. }
 \end{figure}

To explore sensitivity to the surface thickness, we model $\rho_W(r)$ with a two parameter Fermi function $\rho_W(r,c,a)$ \cite{2pf1,2pf2},
\begin{equation}
\rho_W(r,c,a)=\rho_0\frac{{\rm Sinh}(c/a)}{{\rm Cosh}(r/a)+{\rm Cosh}(c/a)}\,.
\label{eq.2pf}
\end{equation}
Here, the parameter $c$ describes the size of the nucleus while $a$ describes the surface thickness (see Table \ref{tab:table1}).  The normalization constant $\rho_0$ is
\begin{equation}
\rho_0=\frac{3Q_W}{4\pi c(c^2+\pi^2 a^2)}\, ,
\end{equation}
so that $\int d^3r \rho_W(r,c,a)=Q_W$.  The $r^2$ and $r^4$ moments of Eq. \ref{eq.2pf} are \cite{2pf1,2pf2},
\begin{equation}
    \langle r^2\rangle =\frac{3}{5}c^2+\frac{7}{5}\pi^2a^2\, ,
\end{equation}
\begin{equation}
    \langle r^4\rangle =\frac{3}{7}c^4+\frac{18}{7}c^2\pi^2a^2+\frac{31}{7}\pi^4a^4\, .
\end{equation}
For the FSU Gold relativistic mean field interaction \cite{FSUGold}, or other density functional, we calculate the $r^2$ and $r^4$ moments and invert the above Eqs. to determine fit parameters $c$ and $a$. 
The results for $^{208}$Pb are listed in Table \ref{tab:table1} and shown in Fig. \ref{fig:2}.  The Fermi function provides a good fit to $\rho_W(r)$ and averages over the small interior shell oscillations.

\begin{figure}[hbt]
\includegraphics[scale=0.3]{{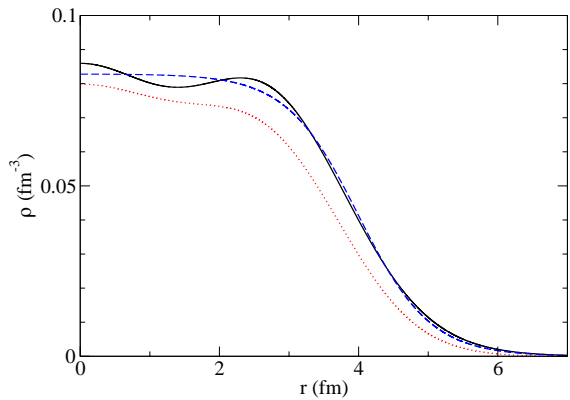}}
\caption{\label{fig:2b} As Fig. \ref{fig:2} except for $^{48}$Ca.}
\end{figure}

The Fermi function fit for $^{48}$Ca is shown in Fig. \ref{fig:2b}: in this case the fit is less good because the interior shell oscillations have larger amplitudes.  Nevertheless, the Fermi function still provides a good qualitative description of $\rho_W(r)$ and we expect the interior shell oscillations to only be important at larger momentum transfers.  Therefore the Fermi function is adequate for our purposes of providing a simple representation of the surface thickness.  Of course, our choice of Eq. \ref{eq.2pf} introduces some model dependence into our analysis.  However, we expect this to be small and other representations of the weak density such as using a Helm form \cite{helm} should give very similar results.  We explore in Sec. \ref{sec.results} the sensitivities of the PREX-II and CREX experiments to values of the surface parameter $a$.

\section{Sensitivity of PREX-II and CREX to the surface thickness}
\label{sec.results}

\begin{figure}[hbt]
\includegraphics[scale=0.3]{{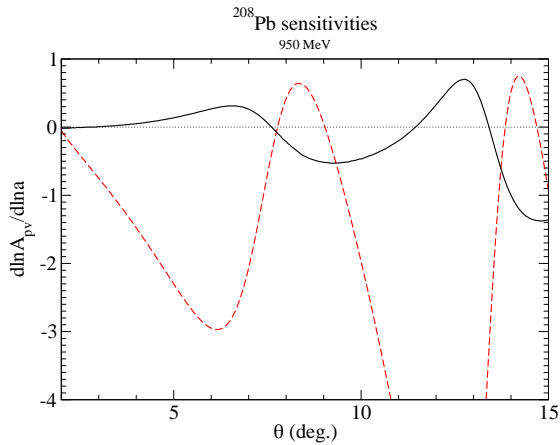}}
\caption{\label{fig:4} Log derivative of the parity violating asymmetry w.r.t. the log of the surface thickness parameter $a$ (solid black line) or w.r.t the log of the weak radius $R_W$ (dashed red line) for $^{208}$Pb at 950 MeV versus scattering angle. }
 \end{figure}

In this section we calculate how $A_{pv}$ depends on the surface thickness parameter $a$ for the kinematics of the PREX-II \cite{PREXII} and CREX \cite{CREX} experiments.  The PREX-II experiment aims to measure the radius $R_W$ of the weak charge distribution of $^{208}$Pb,  
\begin{equation}
R_W^2=\frac{1}{Q_W}\int d^3r\, r^2\, \rho_W(r)=\frac{3}{5}c^2+\frac{7}{5}\pi^2a^2\, .
\label{eq.rw}
\end{equation}
Here the second relation follows from our Fermi function in Eq. \ref{eq.2pf}.  To calculate the sensitivity to changes in the surface thickness $a$ we also change $c$ in such a way that $R_W$ in Eq. \ref{eq.rw} remains constant.  We define the sensitivity to the surface thickness $\epsilon_a$ as the log derivative of the asymmetry w.r.t. the log of the surface thickness,
\begin{equation}
\epsilon_a=\frac{d{\rm ln}A_{pv}}{d{\rm ln}a}=\frac{a}{A_{pv}}\frac{dA_{pv}}{da}\, .
\label{eq.ea}
\end{equation}
We approximate this as $\epsilon_a\approx 100 \frac{1}{A_{pv}} \Delta A$ where $\Delta A$ is the difference in $A_{pv}$ calculated with $a$ increased by 1\% (at constant $R_W$) and the original $A_{pv}$.  This is shown in Fig. \ref{fig:4} for $^{208}$Pb at a beam energy of 950 MeV.  We see that $\epsilon_a$ is very small at angles $<4^\circ$ because at forward angles one is most sensitive to $R_W$ and this has been kept fixed.

We also define the sensitivity to $R_W$ as the log derivative of $A_{pv}$ w.r.t. the log of $R_W$,
\begin{equation}
\epsilon_R=\frac{d{\rm ln}A_{pv}}{d{\rm ln}R_W}=\frac{R_W}{A_{pv}}\frac{dA_{pv}}{dR_W}\, .
\label{eq.eR}
\end{equation}
We approximate this as $\epsilon_R\approx 100 \frac{1}{A_{pv}} \Delta A$ where $\Delta A$ is the difference in $A_{pv}$ calculated with both $c$ and $a$ increased by 1\% and the original $A_{pv}$.  This is also shown in Fig. \ref{fig:4}.  Both $\epsilon_a$ and $\epsilon_R$ are seen to change sign near the first diffraction minimum in the cross section, see Fig. \ref{fig:1}.

At the PREX-II average kinematics 950 MeV and scattering angle $\theta\approx 4.55^\circ$ we find $\epsilon_a\approx 0.091$.  For example, if $A_{pv}$ is measured to about 2.5\% then one will be sensitive to $a$ to $\pm2.5\%/0.091$ or $\pm 27$\%.   If $a$ is known to better than 27\%, the uncertainty in $a$ will not give a large error in the extraction of $R_W$.   We conclude, PREX-II is not very sensitive to the surface thickness $a$.  This is in agreement with previous work, see for example \cite{furnstahl}. 

\begin{figure}[hbt]
\includegraphics[scale=0.3]{{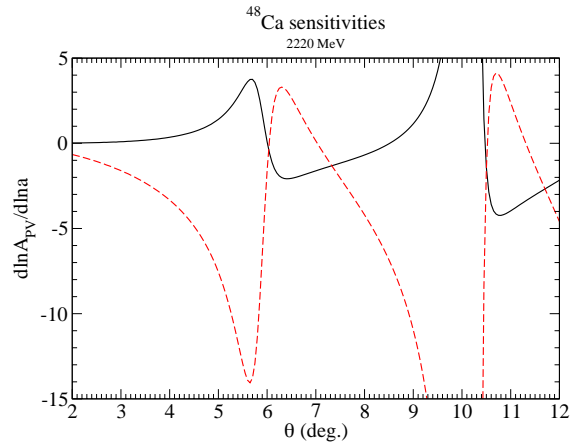}}
\caption{\label{fig:5} As Fig. \ref{fig:4} except for $^{48}$Ca at 2220 MeV. }
\end{figure}

We repeat these calculations for CREX, see Fig. \ref{fig:5}.  At the CREX average kinematics 2220 MeV and $\theta\approx 4.55^\circ$ we find a much large $\epsilon_a\approx 0.731$.  This larger value is because $^{48}$Ca has a larger surface to volume ratio than $^{208}$Pb and because the CREX kinematics are closer to the diffraction minimum.  If $A_{pv}$ is measured to $\approx 5$\%, one will be sensitive to $a$ to $5\%/0.73\approx 7$\%.  Therefore, unless $a$ is known to significantly better than 7\% (which may be unlikely), its uncertainty will be important in the extraction of $R_W$ from $A_{pv}$.  We conclude that CREX is sensitive to the surface thickness and one must carefully address this sensitivity in any extraction of $R_W$.

\section{New experiment to measure the surface thickness}
\label{sec.additional}
In this section we evaluate the statistical error and figure of merit (FOM) for a new parity violating electron scattering experiment to determine the surface thickness of the weak charge density of either $^{208}$Pb or $^{48}$Ca.  The surface thickness of $\rho_W(r)$ can differ from the known surface thickness of $\rho_{ch}(r)$ and is expected to be sensitive to poorly constrained isovector gradient terms in energy functionals.  One way to constrain these gradient terms is to perform microscopic calculations of pure neutron drops in artificial external potentials, using two and three neutron forces. Then one can fit the resulting energies and neutron density distributions with an energy functional by adjusting the isovector gradient terms.  It may be possible to test these theoretically constrained isovector gradient terms by measuring the surface behavior of $\rho_W(r)$.

For reference, we collect in Table \ref{tab:table2} surface thickness parameters $a$ from Fermi function fits to the weak charge density as predicted by a small selection of non-relativistic and relativistic mean field models.  The results of these models  for $^{208}$Pb have an average value of $a\approx 0.60$ fm with a variance of $\pm 0.02$ fm, a $\pm3\%$ range.  For $^{48}$Ca the models give $a\approx 0.525\pm 0.025$ fm, a $\pm5\%$ range.  The models that we have chosen provide some examples.  We caution that these numbers are model dependent.  It is important to measure $a$ in a new parity violating experiment to have a model independent determination.   

\begin{table}[hbt]
\begin{ruledtabular}
\begin{tabular}{ccc}
 Model &  $a[^{208}$Pb] (fm)   & $a[^{48}$Ca] (fm)  \\
\hline
SIII  &  0.5792   & 0.5053 \\
SLY4 & 0.6040 & 0.5247  \\
SV-min & 0.6056 & 0.5386 \\
TOV-min & 0.6111 & 0.5435 \\
UNEDF0 & 0.6155 & 0.5458 \\
IUFSU & 0.6079 & 0.5298 \\
FSU Garnet & 0.6106 & 0.5264 \\
NL3 & 0.6096 & 0.5235 \\
\end{tabular}
\end{ruledtabular}
\caption{\label{tab:table2} Surface thickness parameter $a$ of Fermi function fits to the weak charge densities of $^{208}$Pb and $^{48}$Ca,  see Eq.~\ref{eq.2pf}.}
\end{table}

\begin{figure}[hbt]
\includegraphics[scale=0.3]{{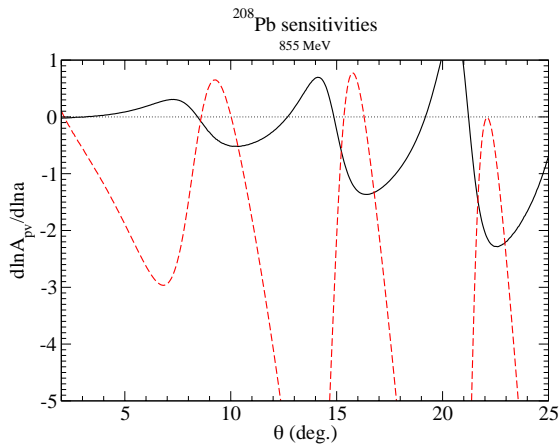}}
\caption{\label{fig:7} As Fig. \ref{fig:4} except for $^{208}$Pb at 855 MeV. }
 \end{figure} 
 
\begin{figure}[hbt]
\includegraphics[scale=0.3]{{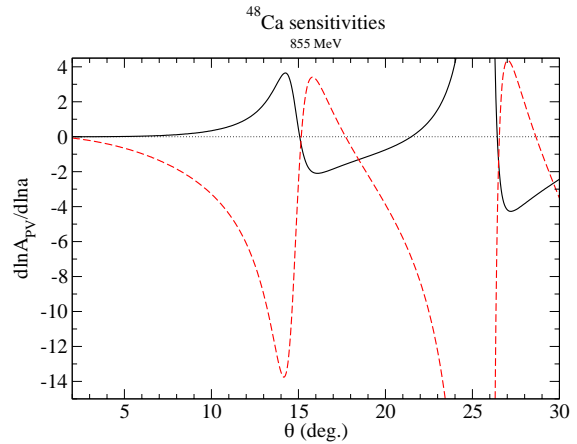}}
\caption{\label{fig:8} As Fig. \ref{fig:4} except for $^{48}$Ca at 855 MeV. }
\end{figure}

%\begin{table}
%\begin{ruledtabular}
%\begin{tabular}{cc}
%     Parameter Value\\
%\hline
%$I$ & 150$\mu$A\\
%$P$ & 0.9 \\
%$\rho_{\rm tar}$ & $2.4 \times 10^{22}$ cm$^{-2}$\\
%$\Delta\Omega$ & 0.0037 Sr \\
%N & 2\\
%$\zeta$ & 0.34\\
%\end{tabular}
%\end{ruledtabular}
%\caption{\label{tab:table2}Assumed experimental parameters including beam current
%$I$, beam polarization $P$, target thickness $\rho_{\rm tar}$, detector solid angle $\Delta\Omega$, number of
%arms N, and radiation loss factor $\zeta$. }
%\end{table}

We now examine the optimal kinematics for an experiment.  The total number of electrons $N_{tot}$ that are scattered into a solid angle $\Delta\Omega$ in a measurement time $T$ is, 
\begin{equation}
N_{tot}=IT\rho_{tar}\frac{d\sigma}{d\Omega}\Delta\Omega\, .
\label{eq:8}
\end{equation}
Here $I$ is the beam current and $\rho_{tar}$ is the density of the target in atoms per cm$^2$.  For simplicity, we neglect radiative corrections.  The statistical error in the determination of the surface thickness $a$ is $\Delta a$,
\begin{equation}
\frac{\Delta a}{a}=\Bigl(N_{tot}A_{pv}^2P^2\epsilon_a^2\Bigr)^{-\frac{1}{2}}\, ,
\label{eq:9}
\end{equation}
where $P$ is the beam polarization.
This depends on the figure of merit ($FOM_a$) that we define as,
\begin{equation}
FOM_a=\frac{d\sigma}{d\Omega} A_{pv}^2 \epsilon_a^2\, .
\end{equation}
One can adjust the scattering angle (or momentum transfer) to maximize $FOM_a$.  This in turn will minimize the run time necessary to achieve a given statistical error in the determination of $a$.  Likewise, the statistical error in the determination of the weak radius $R_W$ is,
\begin{equation}
\frac{\Delta R_W}{R_W} = \Bigl(N_{tot}A_{pv}^2P^2\epsilon_R^2\Bigr)^{-\frac{1}{2}}\, .
\label{eq.FOMa}
\end{equation}
This is closely related to the figure of merit for the determination of $R_W$ that we define as,
\begin{equation}
FOM_R=\frac{d\sigma}{d\Omega} A_{pv}^2 \epsilon_R^2\, .
\label{eq.FOMR}
\end{equation}
The statistical error in the determination of $R_W$ scales with $(FOM_R)^{-1/2}$.

\begin{figure}[hbt]
\includegraphics[scale=0.3]{{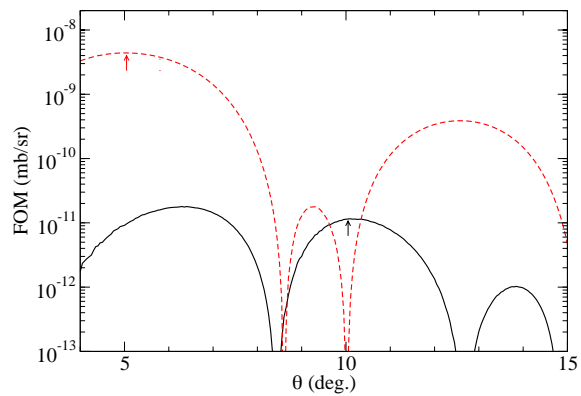}}
\caption{\label{fig:9} Figures of merit $FOM_a$ and $FOM_R$ versus scattering angle $\theta$ to measure the weak surface thickness (solid black curve) or radius (dashed red curve) of $^{208}$Pb at 855 MeV.}
\end{figure} 

\begin{figure}[hbt]
\includegraphics[scale=0.3]{{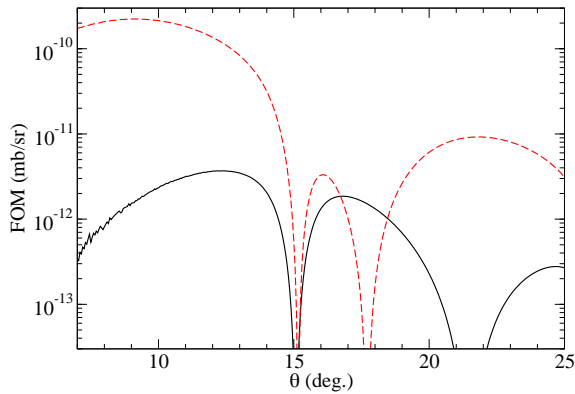}}
\caption{\label{fig:10} Figures of merit $FOM_a$ and $FOM_R$ versus scattering angle $\theta$ to measure the weak surface thickness (solid black curve) or radius (dashed red curve) of $^{48}$Ca at 855 MeV.}
\end{figure} 

%\begin{figure}[hbt]
%\includegraphics[scale=0.3]{{FOMCa570.pdf}}
%\caption{\label{fig:11} Figure of merit $(d\sigma/d\Omega) A_{pv}^2\epsilon_a^2$ versus scattering angle $\Theta$ to measure the weak surface thickness (solid black curve) or radius (dashed red curve) of $^{48}$Ca at 570 MeV.}
%\end{figure} 

In Figs.~\ref{fig:7} and \ref{fig:8} we plot sensitivities $\epsilon_a$ and $\epsilon_R$ for $^{208}$Pb and $^{48}$Ca and we show the figure of merits $FOM_a$ and $FOM_R$ in Figs.~\ref{fig:9} and \ref{fig:10}.  At a laboratory energy of 855 MeV, the maximum in $FOM_R$ for $^{208}$Pb occurs near a scattering angle of 5 degrees.  This is indicated by a red arrow in Fig.~\ref{fig:9} and corresponds to a momentum transfer $q=0.38$~fm$^{-1}$, see Table~\ref{tab:table3}.  A maximum in $FOM_a$ for $^{208}$Pb occurs near 10 degrees or $q=0.76$~fm$^{-1}$.  A parity experiment near this momentum transfer will be sensitive to the surface thickness $a$.   Note there is also a local maximum in $FOM_a$ near 6.5 degrees (or $q=0.49$ fm$^{-1}$) in Fig. \ref{fig:9}.  However a measurement of $A_{pv}$ at this $q$ may be linearly dependent with the PREX-II measurement at only slightly smaller $q\approx 0.38$ fm$^{-1}$.  This could make it difficult to separately determine both $R_W$ and $a$.

%%%%%%%%%%%%%%%%%%%%%%%%%%%%%%%%%%%%%%%%%%%%%%%%%%%%%%%%%%%%
\begin{table}[hbt]
\begin{ruledtabular}
\begin{tabular}{ccc}
Parameter &  $^{208}$Pb   & $^{48}$Ca \\
\hline
$R_W$ & 0.38~fm$^{-1}$ &0.68~fm$^{-1}$ \\
$a$  &  0.76~fm$^{-1}$ & 1.28~fm$^{-1}$ \\
\end{tabular}
\end{ruledtabular}
\caption{\label{tab:table3} Near optimal momentum transfer $q$ that gives large figures of merit, for measuring the weak radius $R_W$ or surface thickness $a$ for the nuclei $^{208}$Pb and $^{48}$Ca.}
\end{table}
%%%%%%%%%%%%%%%%%%%%%%%%%%%%%%%%%%%%%%%%%%%%%%%%%%%%%%%%%%%%

We now scale our results in Figs.~\ref{fig:9} and \ref{fig:10} to other energies.  As long as the energy is above say 400 MeV, Coulomb distortions do not depend strongly on energy.  As a result $A_{pv}$ depends primarily on momentum transfer $q$ and only weakly on beam energy $E$,
\begin{equation}
A_{pv}(E_1,q)\approx A_{pv}(E_2,q)\, .
\end{equation} 
Therefore $\epsilon_i(q)$ in Fig. \ref{fig:5}, as a function of $q$, is very close to $\epsilon_i(q)$ in Fig. \ref{fig:8}.
The differential cross section depends strongly on $q$.  However at fixed $q$, it scales approximately as $E^2$ so that $\frac{d\sigma}{d\Omega}(E_1,q)\approx \frac{E_1^2}{E_2^2}\, \frac{d\sigma}{d\Omega}(E_2,q)$.  Therefore $FOM_i$, at fixed $q$, grows with increasing energy,
\begin{equation}
FOM_i(E_1,q)\approx \frac{E_1^2}{E_2^2}\, FOM_i(E_2,q)\, .
\end{equation}
This is true for both $i=R$ and $a$.  If the solid angle of the detector $\Delta\Omega$ is fixed, it can be advantageous to measure at as forward an angle as possible, and at a higher beam energy, because this will increase the figure of merit.

A measurement of $A_{pv}$ for $^{208}$Pb at $q=0.76$~fm$^{-1}$, see Table \ref{tab:table3}, is sensitive to $a$.  In general, it will also be sensitive to $R_W$.  However PREX-II and CREX are constraining $R_W$ for both $^{208}$Pb and $^{48}$Ca.  Therefore, it should be possible to extract $a$ from only a single new measurement.  This would completely determine a Fermi function model of the weak charge density. 

The statistical error in the extraction of $a$ scales with $FOM_a^{-1/2}$.  For $^{208}$Pb, the maximum in $FOM_a$ in Fig.~\ref{fig:9} near 10 degrees is about 400 times smaller than the maximum in $FOM_R$.  Therefore $a$ can be determined to 10\% with only a few times larger integrated luminosity (beam time) than PREX-II is using to determine $R_W$ to approximately 1\%.  For $^{48}$Ca the local maximum in $FOM_a$ in Fig.~\ref{fig:10} near 17 degrees is only $\approx 1/100$ the maximum in $FOM_R$.  Therefore $a$ can be determined to 10\% using comparable integrated luminosity (or beam time) as a 1\% measurement of $R_W$.   We conclude that a parity violating electron scattering experiment to measure $a$ is feasible.

%tt Add short summary of future MAMI experiment
A similar experiment will be possible at the A1 spectrometer facility of the MAMI accelerator.
According to the construction of the spectrometers and their arrangement on the pivot surrounding the scattering chamber, there are limitations in the accessible angular range. Furthermore, only selected beam energies at MAMI are equipped with a special stabilization system which is essential for performing parity-violation experiments. Two scenarios are currently under investigation, at beam energies of 855 and 570 MeV respectively, to optimise the experimental conditions. 
Cherenkov detectors specifically designed for counting experiments \cite{esser18} will be placed in the focal planes of the high resolution spectrometers. This set-up will allow to make use of the high precision tracking detectors of the spectrometer to align the elastic line of $^{208}$Pb with the Cherenkov detectors by changing the magnetic field setting.
The data acquisition electronics will be the one of the former A4 experiment \cite{a409}.
With a beam intensity of 20 $\mu$A at a scattering angle of 10.35 a 10\% measurement of the surface thickness will be possible in about 100 days for a beam energy of 855 MeV. At the lower beam energy the precision in the extraction of the surface radius will be the same for a scattering angle of 15.2 but the total running time will double.

\section{Conclusions}
\label{sec.conclusion}
The present PREX-II and CREX parity violating electron scattering experiments are probing the weak charge densities of $^{208}$Pb and $^{48}$Ca.  These experiments are primarily sensitive to the radius of the weak charge density $R_W$ but they are also sensitive to the surface thickness $a$.  In this paper we have explored the feasibility of a new parity violating electron scattering experiment to measure $a$ for a neutron rich nucleus.  PREX-II or CREX combined with an additional parity violating measurement at a momentum transfer near 0.76 fm$^{-1}$ for $^{208}$Pb or 1.28 fm$^{-1}$ for $^{48}$Ca will cleanly determine both $R_W$ and $a$.  Determining $a$ both sharpens the determination of $R_W$ from PREX II or CREX and determines the average interior weak charge density and baryon density \cite{nuclearsaturation}.  In particular, the average interior baryon density of $^{208}$Pb is closely related to the saturation density of nuclear matter.

%The ground state neutron density of a medium mass nucleus contains fundamental nuclear structure information and it is at present relatively poorly known.  In this paper we explored if parity violating elastic electron scattering can determine not just the neutron radius, but the entire radial form of the neutron density $\rho_n(r)$ or weak charge density $\rho_W(r)$ in a model independent way.  We expanded the weak charge density $\rho_W(r)$ in a model independent Fourier Bessel series.  For the medium mass neutron rich nucleus $^{48}$Ca, we find that a practical parity violating experiment could determine about six Fourier Bessel coefficients $a_i$ and thus deduce the full radial structure of both $\rho_W(r)$ and the neutron density $\rho_n(r)$.  The resulting $\rho_W(r)$ will contain fundamental information on the size, surface thickness, shell oscillations, and saturation density of the neutron distribution.

%Future work could optimize our model experiment to further reduce the statistical errors by for example using large acceptance detectors and combining information from multiple experiments and or laboratories.   Future theoretical work exploring the range of weak charge densities to be expected with reasonable models and microscopic calculations would also be very useful.  The measured $\rho_W(r)$, combined with the previously known charge density $\rho_{ch}(r)$, will literally provide a detailed textbook picture of where the neutrons and protons are located in an atomic nucleus. 

\section*{Acknowledgements}

%We thank Bob Michaels for helpful comments and Shufang Ban for initial contributions to this work.  We thank the Mainz Institute for Theoretical Physics for their hospitality.  
This research was supported in part by the United States Department of Energy Office of Science, Office of Nuclear Physics grants DE-FG02-87ER40365 (Indiana University) and DE-SC0018083 (NUCLEI SciDAC Collaboration).

   % input acknowledgement

\end{document}